\documentclass[
    ,final            % use final for the camera ready runs
  ]
  {aipproc}

\layoutstyle{6x9}
\usepackage{natbib}

\begin{document}

\title{New optical polarization measurements of the Crab pulsar}


\author{G. Kanbach}{
  address={Max Planck Institute for Extraterrestrial Physics, Postfach 1312,
D-85741 Garching, Germany}
}

\author{A. S{\l}owikowska}{
  address={Nicolaus Copernicus Astronomical Center, Rabia\'nska 8, 87-100 Toru\'n, Poland}
}

\author{S. Kellner}{
  address={Max Planck Institute for Astronomy, K\"onigstuhl 17,
D-69117 Heidelberg, Germany}
}

\author{H. Steinle}{
  address={Max Planck Institute for Extraterrestrial Physics, Postfach 1312,
D-85741 Garching, Germany}
}

\begin{abstract}
The Crab nebula and its pulsar have been observed
for about 3 hours with the high-speed
photo-polarimeter OPTIMA in January 2002 at the Calar Alto 3.5~m telescope. The Crab
pulsar intensity and polarization are determined at all phases 
of rotation with higher statistical accuracy than ever.
Therefore, we were able to separate the 
so-called 'off-pulse' phase emission (with an intensity of about 1.2\% compared
to the main peak, assumed to be present at all phases) from the pulsed emission
and show the 'net' polarization of the pulsed structures. Recent theoretical
results indicate that the measured optical polarization of the Crab pulsar 
is similar to expectations from a two-pole caustic emission model or a striped 
pulsar wind model.
\end{abstract}

\maketitle

%%%%%%%%%%%%%%%%%%%%%%%%%%%%%%%%%%%%%%%%%%%%
%% MAINMATTER
%%%%%%%%%%%%%%%%%%%%%%%%%%%%%%%%%%%%%%%%%%%%
\section{Instrument and Observations}
OPTIMA (Optical Pulsar TIMing Analyzer) has been built and developed at
Max Planck Institute for Extraterrestrial Physics in Garching. The instrumental 
sensitivity (white light) extends from  about 450~nm to 950~nm and the used polaroid
filter modulates the incoming radiation effectively for wavelengths shorter
than 850~nm. Astronomical targets are imaged onto a hexagonal bundle of optical 
fibers which are coupled  to single avalanche photodiode photon counters. 
The spacing and size of 
the fibers corresponds to about 2 arcsec. GPS based time tagging of single photons,
together with the instantaneous determination of the orientation of a rotating
polaroid filter (by using the Hall probe), allows to measure the phase
dependent linear polarization state of the pulsar and the surrounding nebula
simultaneously. Detailed description of the instrument 
can be found at the MPE OPTIMA web page 
\footnote{\texttt{http://www.mpe.mpg.de/gamma/instruments/optima/www/optima.html}},
as well as in \cite{Straubmeier2001}, \cite{Kellner2002}, and \cite{Kanbach2003}.
The observations were performed in January 2002 by using the OPTIMA instrument
attached to the 3.5~m telescope  at the Calar Alto Observatory.
\section{Polarization of total emission of the Crab pulsar}
%%% Fig. 1
\begin{figure}
  \includegraphics[height=.45\textheight]{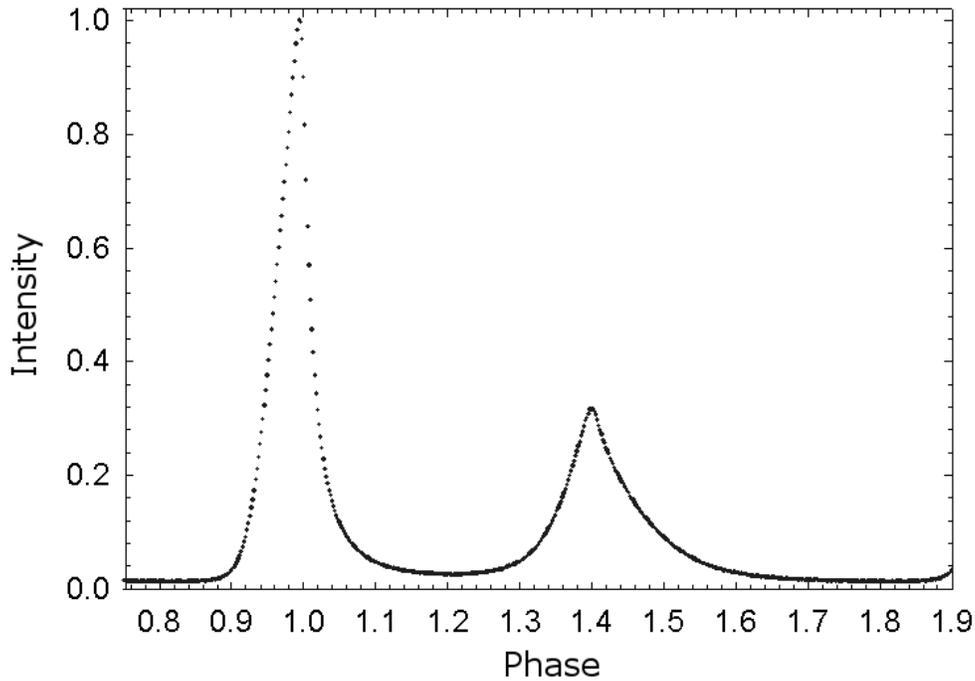}
  \caption{Normalized light curve of the Crab pulsar
obtained by OPTIMA at Calar Alto Observatory in January 2002.
The so-called 'off-pulse' phase emission within phase range
1.72-1.82 is on the level of 1.2\% of the main peak intensity.}
  \label{lc}
\end{figure}
\begin{figure}
  \includegraphics[height=.65\textheight]{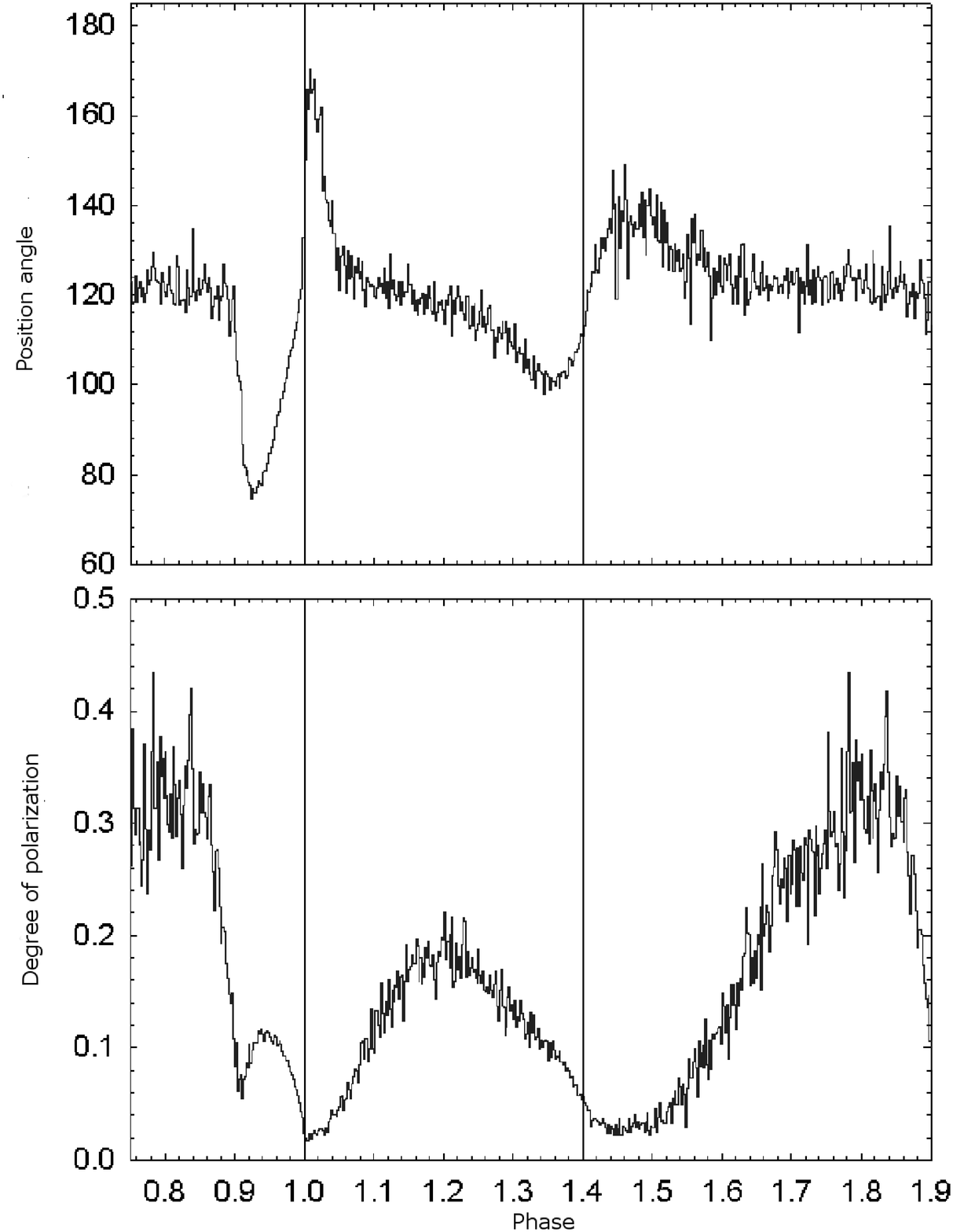}
  \caption{The position angle (E by N) 
and the degree of polarization of the
E-vector on the sky for the Crab pulsar. This result are generally
consistent with a previous measurement by \cite{Smith1988}
but shows details with much better definition and statistics.
Please note that the degree of polarization
has to be considered as preliminary. A correction for the wavelength
dependent modulation depth of  the polaroid filter used in this
measurement has not yet been applied. We expect the degree of
polarization in the corrected version to be slightly larger than
shown here.}
  \label{polall}
\end{figure}

The Crab pulsar is detected at all phases of
rotation (Fig.~\ref{lc}), i.e. also in the so-called 'off-pulse'
phase with an intensity of about 1.2\% compared to the intensity of
the main peak. This confirms the result of \cite{Golden2000}.
From this measurement the polarization characteristics of the Crab
pulsar shown in Fig.~\ref{polall} were derived.
The degree of polarization and the position angle (P.A.) of the E-vector
are plotted with a resolution of 500 phase bins.  Our result agrees
generally well with the previous measurements (\cite{Smith1988}), 
but shows details with much better definition and statistics. 
The variations of the P.A. observed for the Crab at optical wavelengths differ from those
observed at radio wavelengths (\cite{Moffett1999}, \cite{Kar2004}, 
\cite{Slowikowska2005}). There might be two reasons for it.
Firstly, it can be caused by the restricted range of altitudes 
where the radio emission at any fixed frequency originates
(contrary to the wide range of altitudes assumed in the high
energy emission models, e.g. TPC=two pole caustic and OG=outer gap).
Secondly, the P.A. may depend strongly on photon energy.
%%% Fig. 2

\section{Polarization of the pulsed component only}
\begin{figure}
  \includegraphics[height=.75\textheight]{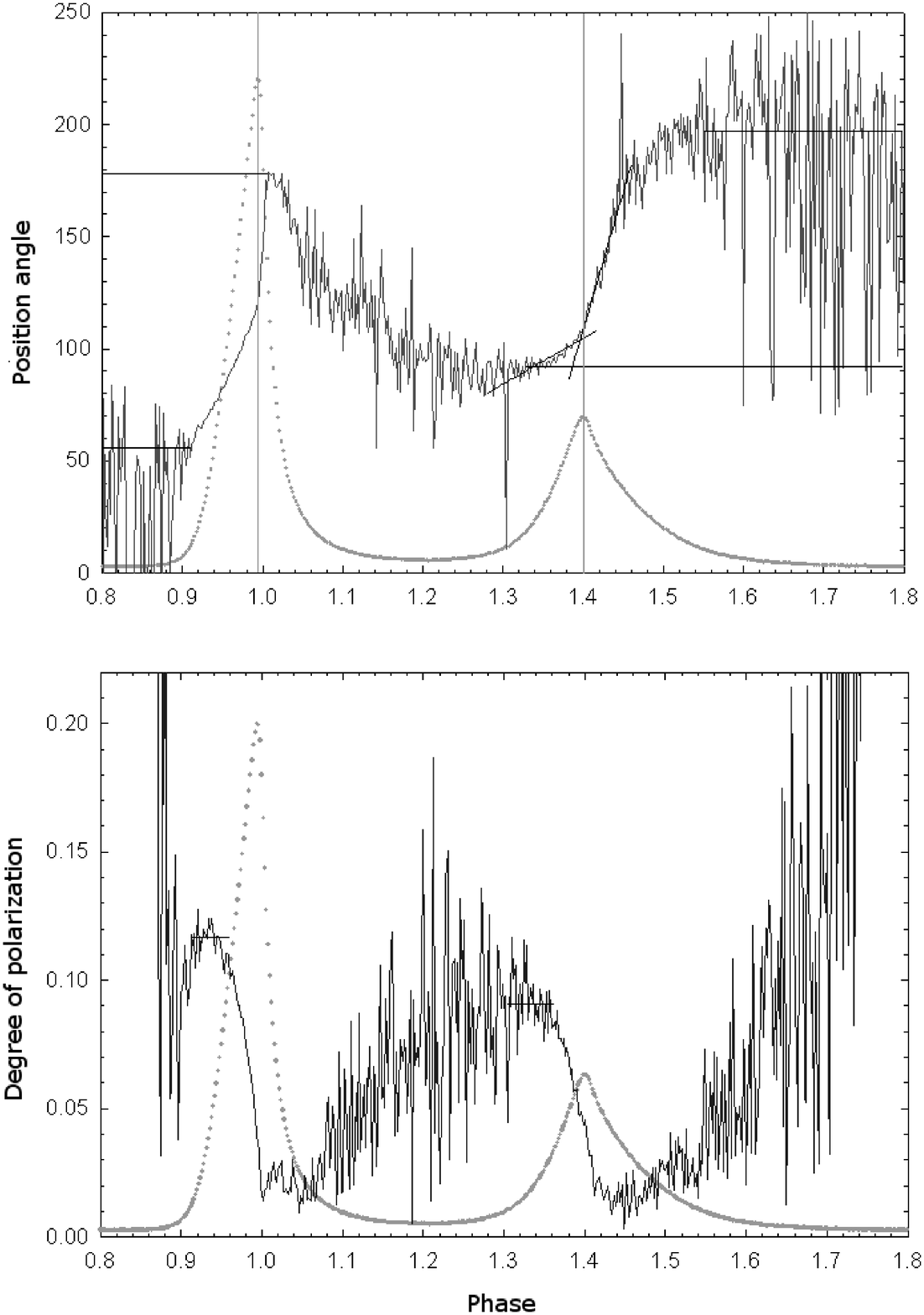}
  \caption{Polarization characteristics, P.A. and the degree of
polarization,  of the 'pulsed component' of the Crab pulsar.
For clarity the light curve of the Crab pulsar is overplotted
(grey dotted-line).}
  \label{polpulsed}
\end{figure}

\begin{figure}
  \includegraphics[height=.55\textheight]{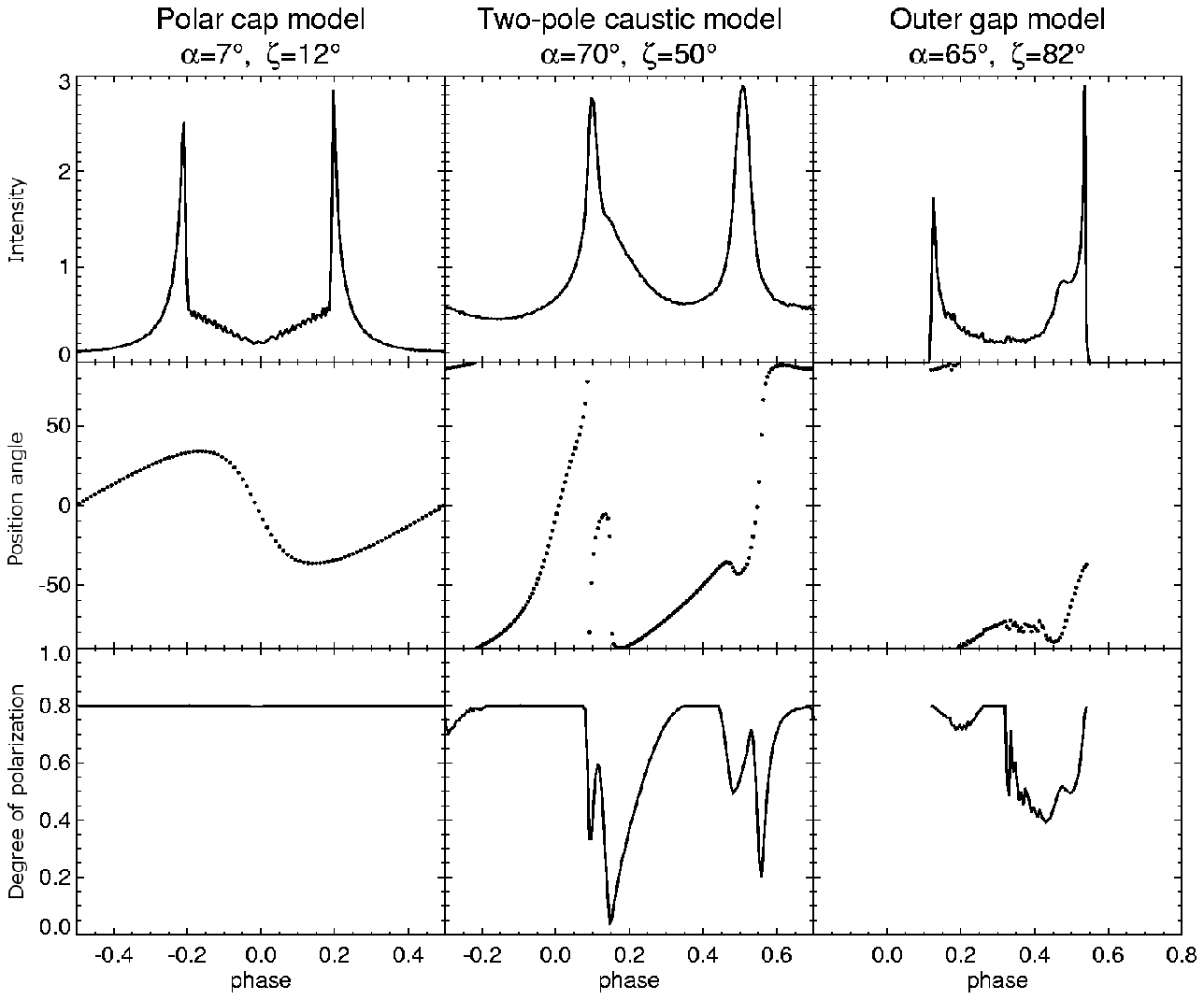}
  \caption{The optical light curve, the position angle and the degree of
polarization calculated with the following models of high energy radiation
from pulsars, from left to right:  the polar cap model,
the two-pole caustic model, and the outher gap model
(\cite{DyksIAUS2004}).}
  \label{model}
\end{figure}

Constancy of the position angle within the phase range 1.7-1.9
(Fig.~\ref{polall}) may suggest that the optical emission from the Crab
pulsar consists of two, pulsed and unpulsed, components: (i) one
characterized by a highly variable P.A. and polarization degree, and (ii)
a DC component with constant intensity on the level of 1.24\% of the main pulse
intensity, fixed P.A. (123$^\circ$), and a degree of polarization on the
level of 33\%.  Assumming that the continous component is present at all
phase angles and has constant polarization we obtained the polarization
characteristics of the 'pulsed component' separately (Fig.~\ref{polpulsed})
by subtracting the respective Stokes parameters I, Q, U
from the linear polarization data. For comparison
the light curves and polarization characteristics obtained within the
framework of three high energy magnetospheric emission models of 
pulsars, the polar cap model, the two-pole caustic model, 
and the outer gap model are shown in Fig.~\ref{model}.
%%% Fig. 3
%%% Fig. 4
The two-pole caustic model (\cite{Dyks2003})
predicts fast swings of the postion angle
and minima in the polarization degree, similar to what is observed.
Polar cap model and outer gap model are not able to reproduce the 
observational polarization characteristics of the Crab pulsar. 
Another model, placeing the origin of the pulsed optical emission from the Crab
in a striped pulsar wind zone has recently been proposed by \cite*{Petri2005}.
This model features also polarization characteristics that bear a certain 
resemblance to the observations.

\section{Conclusions}
\begin{itemize}
\item
The optical emission from the Crab pulsar is highly polarized,
especially in the bridge and 'off-pulse' phase.
\item
The degree and angle of linear polarization show well defined 
structures: 
\begin{itemize}
\item[-]
at the peaks of the light curve  the degree of 
polarization is minimal,
\item[-]
there is a well defined bump in 
polarization on the rising flank of the main pulse
(possibly also for the inter pulse),
\item[-]
the polarization angle swings through a large angle in both peaks: 
after subtraction of an assumed constantly polarized continous emission
(intensity on the level of 1-1.5\% of the main pulse) 
the angles swing for $125^\circ$ and $105^\circ$ for 
the main pulse and inter pulse, respectively.
\end{itemize}
\item
The polarization signatures observed for the Crab show a high 
degree of similarity with theoretical estimates published by \cite*{Dyks2004}
in the framework of the two-pole caustic model of magnetospheric emission or the 
characteristics derived in a striped pulsar wind model by \cite*{Petri2005}. 
\end{itemize}

%%%%%%%%%%%%%%%%%%%%%%%%%%%%%%%%%%%%%%%%%%%%%%%%
%% BACKMATTER
%%%%%%%%%%%%%%%%%%%%%%%%%%%%%%%%%%%%%%%%%%%%%%%%

\begin{theacknowledgments}
This work was supported by KBN grant 2P03D.004.24 (AS).
\end{theacknowledgments}




%\bibliography{kanbach_ref}
%% !!!!!!!!!!!!!!!!!!!!
%% >>> add here J.Petri and J.G. Kirk, ApJ in press (2005), also astro-ph/0505427
%% !!!!!!!!!!!!!!!!!!!!

\IfFileExists{\jobname.bbl}{}
 {\typeout{}
  \typeout{******************************************}
  \typeout{** Please run "bibtex \jobname" to optain}
  \typeout{** the bibliography and then re-run LaTeX}
  \typeout{** twice to fix the references!}
  \typeout{******************************************}
  \typeout{}
 }

\end{document}